\RequirePackage{fix-cm}

\documentclass[final]{svjour3}                     

\smartqed  
\usepackage{graphicx}
\usepackage{apacite}
\usepackage[english]{babel} 
\usepackage[T1]{fontenc}
\usepackage[utf8]{inputenc}
\usepackage{subfigure}

\usepackage{color} 
\usepackage{soul} 
\usepackage{booktabs} 
\usepackage{latexsym} 
\usepackage[final]{pdfpages}
\usepackage{multirow}
\usepackage{rotating}
\usepackage{setspace}
\usepackage{amsmath}

\usepackage{kpfonts}

\begin{document}
\title{There's more to the multimedia effect than meets the eye: is seeing pictures believing?}
\author{}
\author{Magnus {\"O}gren  \and Marcus Nystr{\"o}m,  \and Halszka Jarodzka}
\institute{M. {\"O}gren \at
              Dep. of Applied Mathematics and Computer Science, Technical University
							of Denmark \&\\
							Nano-Science Center, Department of Chemistry, University of
							Copenhagen, Denmark \&\\
							School of Science and Technology, {\"O}rebro University, SE-701 82
							{\"O}rebro, Sweden. \\
              \email{magnus@ogren.se}           
							\and
							M. Nystr{\"o}m \at
              Helgonabacken 12,  Lund, Sweden \\
              \email{marcus.nystrom@humlab.lu.se}
							\and
							H. Jarodzka \at
              Valkenburgerweg 177, 6419 AT Heerlen, The Netherlands \\
              \email{halszka.jarodzka@ou.nl}
							}  

\date{}

\maketitle
\begin{abstract}
Textbooks in applied mathematics often use graphs to explain the meaning of formulae, even though their benefit is still not fully explored.
To test processes underlying this assumed multimedia effect we collected performance scores, eye movements, and think-aloud protocols from students solving problems in vector calculus with and without graphs. Results showed no overall multimedia effect, but instead an effect to confirm statements that were accompanied by graphs, irrespective of whether these statements were true or false. Eye movement and verbal data shed light on this surprising finding. Students looked proportionally less at the text and the problem statement when a graph was present. Moreover, they experienced more mental effort with the graph, as indicated by more silent pauses in thinking aloud. Hence, students actively processed the graphs. This, however, was not sufficient. Further analysis revealed that
the more students looked at the statement, the better they performed. Thus, in the multimedia condition the graph drew students’ attention and cognitive capacities away from focusing on the statement.
A good alternative strategy in the multimedia condition was to frequently look between graph and problem statement, and thus to integrate their information. In conclusion, graphs influence where students look and what they process, and may even mislead them into believing accompanying information. Thus, teachers and textbook designers should be very critical on when to use graphs and carefully consider how the graphs are integrated with other parts of the problem.
\keywords{multimedia effect \and eye tracking\and verbal data\and mathematics education\and science education}
\end{abstract}

\section*{Introduction}
Mathematical textbooks often include different forms of pictures (such as illustrations, graphs, diagrams, etc.). The reasons are twofold. On the one hand, teachers and textbook designers generally believe that pictures would be helpful for students to better understand 
the material. On the other hand, cognitive theories of information processing  
recommend to enrich scientific texts with pictures to support students in building a rich and coherent mental model of the subject matter (i.e., multimedia effect). However, there are two critical points to this view. First, theories underlying the multimedia effect make statements about perceptual processes that have not been verified directly. Second, recent empirical research questions this general beneficial effect of pictures and even suggests that pictures may bias people into being uncritical towards scientific texts (i.e., picture bias effect) \cite{mccabe2008seeing}.
Thus, in this study, we (1) investigate the multimedia effect and its underlying cognitive and perceptual processes directly with \textit{think-aloud protocols} and \textit{eye tracking} and (2) we test the picture bias effect. Both investigations are in the context of mathematical education at a university level (vector calculus). 

\subsection*{Basic assumptions of learning with multimedia}
Material that presents information in different formats, such as text, pictures, diagrams, and formulae, is referred to as \textit{multimedia}. Two leading theories describe how the human cognitive system processes multimedia material, namely the \textit{Cognitive Theory of Multimedia Learning} (CTML; \citeNP{Ma05}) and the \textit{Cognitive Load Theory} (CLT; \citeNP{SwMePa98}). Both theories assign a central role to working memory \cite{Ba92}. They make three assumptions on the functioning of working memory. 

First, for information to be learned and successfully stored, it has to be \textit{actively processed} in working memory. \citeA{Ma05} describes active processing in three steps: Information has to be selected from a source by means of attention to enter working memory. Next, this information has to be organized into mental models. Last, these mental models have to be integrated with each other and prior knowledge from long-term memory. Only information that has been processed in such a way can be stored in the long-term memory. The `select' and `integrate' processes refer to \textit{perceptual} processes, however, these were only theoretically deduced, but not directly tested.

Second, working memory is of \textit{limited capacity}, which must not be exceeded. \citeA{SwMePa98} proposed that working memory capacity can be filled with three types of loads, namely load caused by active processing of the information (germane load), load stemming from the difficulty of the task (intrinsic load), and by load stemming from other unnecessary cognitive processes that do not contribute to executing the task at hand (extraneous load). The amount of cognitive load posed upon working memory (i.e., mental effort) can be measured with different methods, such as subjective rating scales \cite{Pa92} or silent pauses in thinking aloud \cite{YiCh07,Jarodzka2015}.

Third, both theories assume that \textit{two separate channels} exist for processing verbal and pictorial information \cite{Ba92,Pa86}. 
Both channels are in earlier steps of processing information independent, hence they are loaded with information separately. In later processing steps this information is integrated and leads to a richer mental model than when based on one modality (either pictures or words) only.

Based on these three assumptions both theories (CTL, CTML) provide guidelines on how multimedia material should be designed to optimize cognitive processing of information, as will be described in the next section. It is important to note that both theories and also their resulting guidelines refer to \textit{learning}. Nevertheless, we argue that these theories as well as their resulting guidelines can be adapted to task performance without a specific learning intention, because they are built upon general assumptions of the human cognitive system: First, the assumption that the human cognitive system is limited in capacity with respect to how much information it can process at a time (not in long-term memory storage, though) dates back to early research on the structure and functioning of human working memory \cite<e.g., >{Ba92,Mi56}.
Hence, this assumption holds true not only for learning, but also for general task performance. Second, the active processing assumption of information is based on Atkinson’s and Shiffrin’s information processing model \cite{AtSh68}, which again is not a specific learning model, but instead describes general information processing. Thus, in this study we applied these principles to task performance i.e., solving a problem in vector calculus. 

\subsection*{Guidelines for designing multimedia material: The multimedia principle}
One of the basic guidelines of the above mentioned multimedia theories (CTL, CTML)
is the \textit{multimedia principle}, which assumes that “people learn better from words and pictures than from words alone” \cite{FlTo05, Ma01}. The main idea is that text and pictures evoke different cognitive processes resulting in different mental models which, when later integrated, result in a richer mental model compared to one of the models alone. 
Moreover, when information is presented both in a pictorial and a textual manner, students can use both processing channels in parallel and more efficiently use their working memory. This enables an active processing of information.

A long history of research provides evidence for the multimedia principle (for instance, see the research conducted by the research group of Richard Mayer). \citeA{Ma01a} reports nine of his own studies, all with beneficial learning effects when pictures accompany text.
Confirming this positive effect of pictures, \citeA{CaLe02} present a review with 18 articles from the 90ies reporting beneficial learning-effects of pictures accompanying texts.

In one of his articles, \citeA{Ma89} showed that learning about car mechanics improved when pictures were accompanied with text compared to text only (or pictures only). He explains the multimedia effect by the fact that such illustrations helped students to “focus their attention” and “organize the information into useful mental models”. However, these conclusions were not directly tested.

Thus, both theory and empirical research state that pictures accompanying texts in mathematical problem solving reduce mental effort and help the students to focus their attention, although these assumptions were often deduced from an improved task performance, but not directly tested. 

\subsection*{Limitations and restrictions of the multimedia effect}
Several empirical studies provide challenges for the multimedia principle. 
Often, students do not make use of pictures as was intended. 
For instance, \citeA{BeLi09} found that school children do not benefit from pictures in mathematical problem solving as much as intended. The authors concluded that integrating two information sources probably required more working memory capacity than available 
(for similar findings in school exams see \citeNP{CrSw06}). In line with these findings, \citeA{HoHoHo09} found in a naturalistic newspaper reading study that if pictures and text are given in a standard format, i.e., where they are presented separately in a `split' format, readers often do not make the effort to integrate these information sources (as shown by little visual integration between both information sources indicated with eye tracking). Thus, providing additional information in graphs---irrespective of whether it is relevant to the task---requires additional cognitive resources. If these resources are not available or not allocated correctly graphs can even be harmful. 

\subsubsection*{Bias towards believing}
Other researchers sees the additional use of pictures even more critically. \citeA{LeScMu13}, for instance, showed that pictures reduce the \textit{perceived difficulty} of a learning material. 
This could be very dangerous as students might put too little mental effort into understanding the text so that they do not process all information actively (i.e., select all relevant information from all possible information sources, organize it into coherent mental models, and integrate it), which in turn would result in a poorer task performance. 

Other lines of research unrelated to learning or instruction also critically investigate the effect of pictures. \citeA{IsRiMaKnHoSc13} found that graphs increased the \textit{perceived plausibility} of conflicting information in science text. Again, this is problematic as it could result in students overlooking logical flaws in a text and thus not being able to build a coherent mental model of the task at hand, again, resulting in poorer task performance. 
\citeA{mccabe2008seeing} showed that the mere presence of an illustration increased the \textit{perceived credibility} of a scientific text. 
The readers were less critical against the arguments of a scientific text, when it was accompanied by a scientific illustration. 
As with the other examples, this uncritical attitude towards a text prevents students from building a coherent mental model of its content. Therefore, they are not able to draw the correct conclusions from this mental model, when it has to be applied to perform a particular task.
Hence, pictures that are of a scientific nature may easily be perceived as a proof of the accompanying text and mislead students into believing it---irrespective of whether they do add to its arguments or not. Only a careful integration of both information sources could prevent someone from making this mistake.

\subsection*{Vector calculus as an exemplary mathematical domain}
In the present study the multimedia- and picture bias effects were investigated in the domain of vector calculus. We chose this domain for two reasons. First, vector calculus is a crucial foundation for studies in mathematics and is used in many branches of physics and engineering  \cite<for details on the Swedish curriculum in vector analysis see>{griffiths1999introduction, ramgardvektoranalys, persson1988analys}. 
Second, vector calculus is a very visual topic where an abstract mathematical formula often can be accompanied with a direct graphical representation. 
One of the authors of this article has been teaching courses in vector calculus and has discussed the topic among several colleagues from different countries.
It is a common belief among teachers we have talked to that a key to understanding vector calculus is to be able to switch between different representations of a problem, and successfully integrate the information from all representations into one coherent mental model. This is referring to a deeper form of understanding, necessary for instance to be able to apply relevant knowledge to new applications.

\subsection*{The present study}
\label{sec:present_study}
In this study we investigate whether we can find a general multimedia effect for mathematical problem solving in vector calculus by comparing problem solving tasks with and without accompanying graphs. An example problem is shown in Figure~\ref{fig:stimuli}. Furthermore, we test whether these graphs bias students into believing their accompanying texts by asking students to reject or confirm statements about the task. To better understand the processes underlying the multimedia effect, we use two process-tracing measures: eye tracking \cite{holmqvist2011eye} and verbal reporting \cite{ErSi93}. 
Eye tracking tells us which areas students visually select information from and how they visually integrate these areas. 
Concurrent verbal reporting may provide insight into the amount of mental effort invested by students \cite{YiCh07,Jarodzka2015}. Moreover, it can deliver qualitative information about the underlying processes by serving as a dual-task measure of mental effort \cite<e.g.,>{brunken2003direct,park2015rhythm}.
We hypothesize the following with respect to performance (H$_1$ and H$_2$) and processes (H$_{3a,b,c}$).
\begin{description}
	\item[H$_1$] Performance (i.e., correctly confirming or rejecting a problem statement) is higher with than without graphs, that is, we expect a multimedia effect.
	\item[H$_2$] Confirming the problem statement is more likely with than without graphs, that is, we expect a picture bias effect. 
As a results of the picture bias, we also expect higher performance in the multimedia condition when statements are to be confirmed, compared to the control condition (without a graph).
	\item[H$_3$] Students process information differently depending on whether a graph is present or not. In particular, we expect: 
		\begin{itemize}
			\item[H$_{3a}$]  If a graph is present, students \textit{search and select} information from it. This shows in time spent looking at the graph. 
			Furthermore, as we expect a multimedia effect, we consequently expect that search and selecting information from the graph is positively related to task performance. 
			In addition, we explore to which extent this shift of attention towards the graph happens at the expense of the other information areas (text and formula input \& problem statement) and to which extent attending to these is related to performance; we predict a higher performance the more the graphs are attended.			
			\item[H$_{3b}$] 	If a graph is present, students \textit{integrate} information from it with information from other sources, such as the input (text and formula) and the problem statement. This shows in the amount of transitions between the graph and the other information sources. 
			Furthermore, as we expect a multimedia effect, we consequently expect that integrating from the graph with other sources is positively related to task performance. 
			\item[H$_{3c}$]  In problems with graphs students use more \emph{mental effort} than in problems without graphs, because they need to process more information. This becomes evident in the overall proportion of silence calculated directly from the recorded sound file. A higher proportion of silence is predicted when graphs are present, as a result of the increased mental effort.
		\end{itemize}
\end{description}

Moreover, as an open research question (RQ$_1$), we investigate in two contrasting cases the extent to which  participants follow the processes predicted by the CTMML (search information, build
a mental model, activate prior knowledge, integrate information, and form a problem solution). In addition, we investigate their meta-cognitive and off-topic statements.

\section*{Method}
\subsection*{Participants and design}
Thirty-six students (three females) with an average age of $21.5$ years ($\textit{SD}=3.0$) took part in the experiment. They studied engineering physics (F) at the Lund Institute of Technology (LTH), and were two weeks into a basic course in vector calculus. 
Hence they should be considered as a fairly uniform population with respect to their study background. All students had normal or corrected-to-normal (i.e., with glasses or lenses) vision. They were randomly assigned to one of two conditions in a between-subject design: one solving eight problems without graphs ($N=16$), and one solving the same problems with graphs ($N=20$).

\subsection*{Stimuli}

The stimuli consisted of eight problems dealing with basic concepts in vector calculus. They concerned, for example, simple cases of integration along curves in a two-dimensional domain, the interpretation of the gradient for a function of two variables, and Gauss formula in three dimensions. Each problem was composed of a text and a formula that described a general context, a problem statement that was to be confirmed or rejected and, in the multimedia condition, a graph. 
In this article the word \textit{graph} is used in a broader context than it has in mathematics texts on, e.g., graph theory or functions, which we include, but are not restricted to. In three of the problems, the correct answer was to confirm the problem statement. In the remaining five, a rejection of the statement would provide a correct answer.

\begin{figure}
	\centering
	\subfigure[]{
	\includegraphics[width=0.40\textwidth]{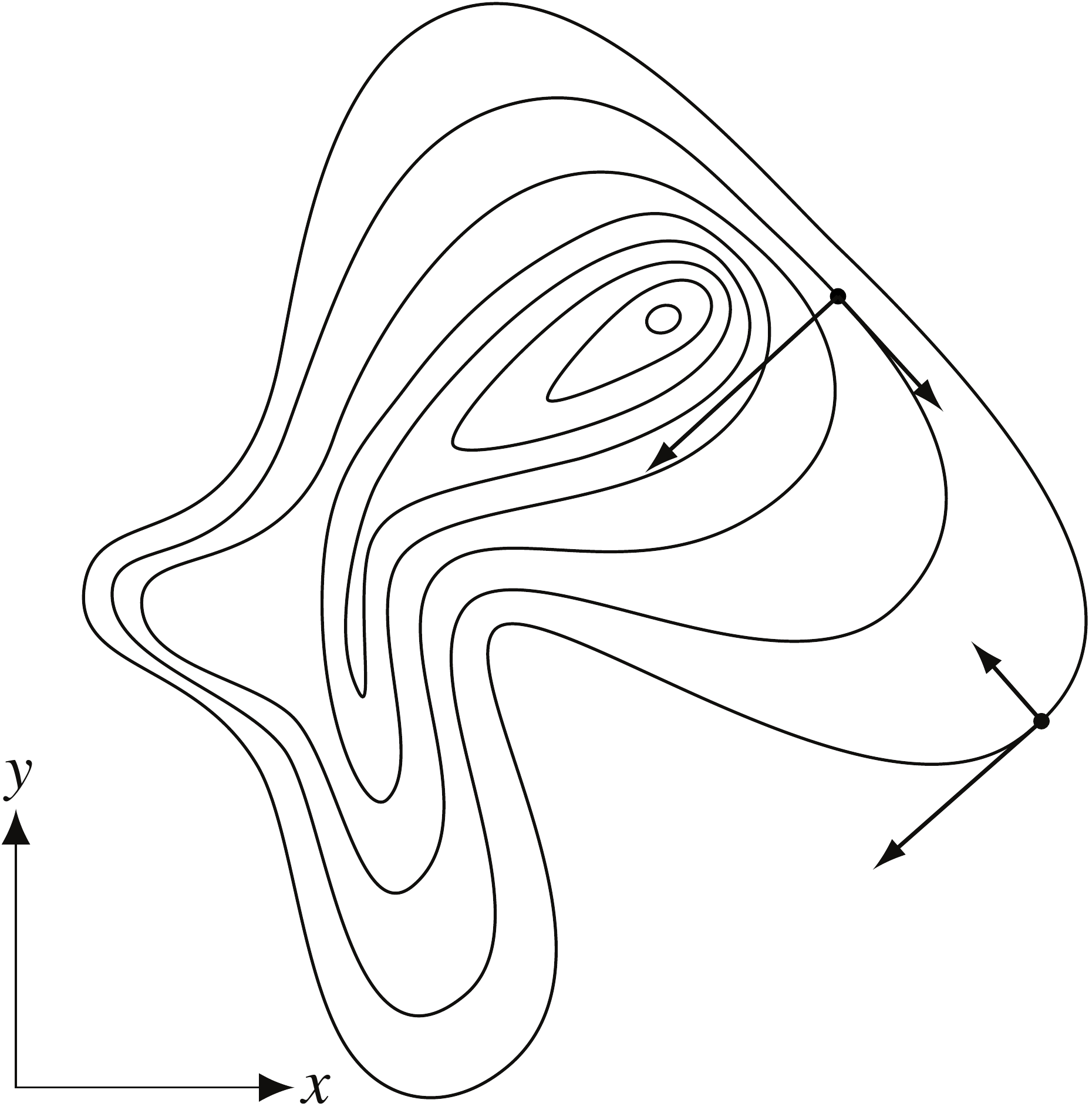}
		\label{fig:fig_Nivakurvor_1_v01}}
	\subfigure[]{\includegraphics[width=0.40\textwidth]{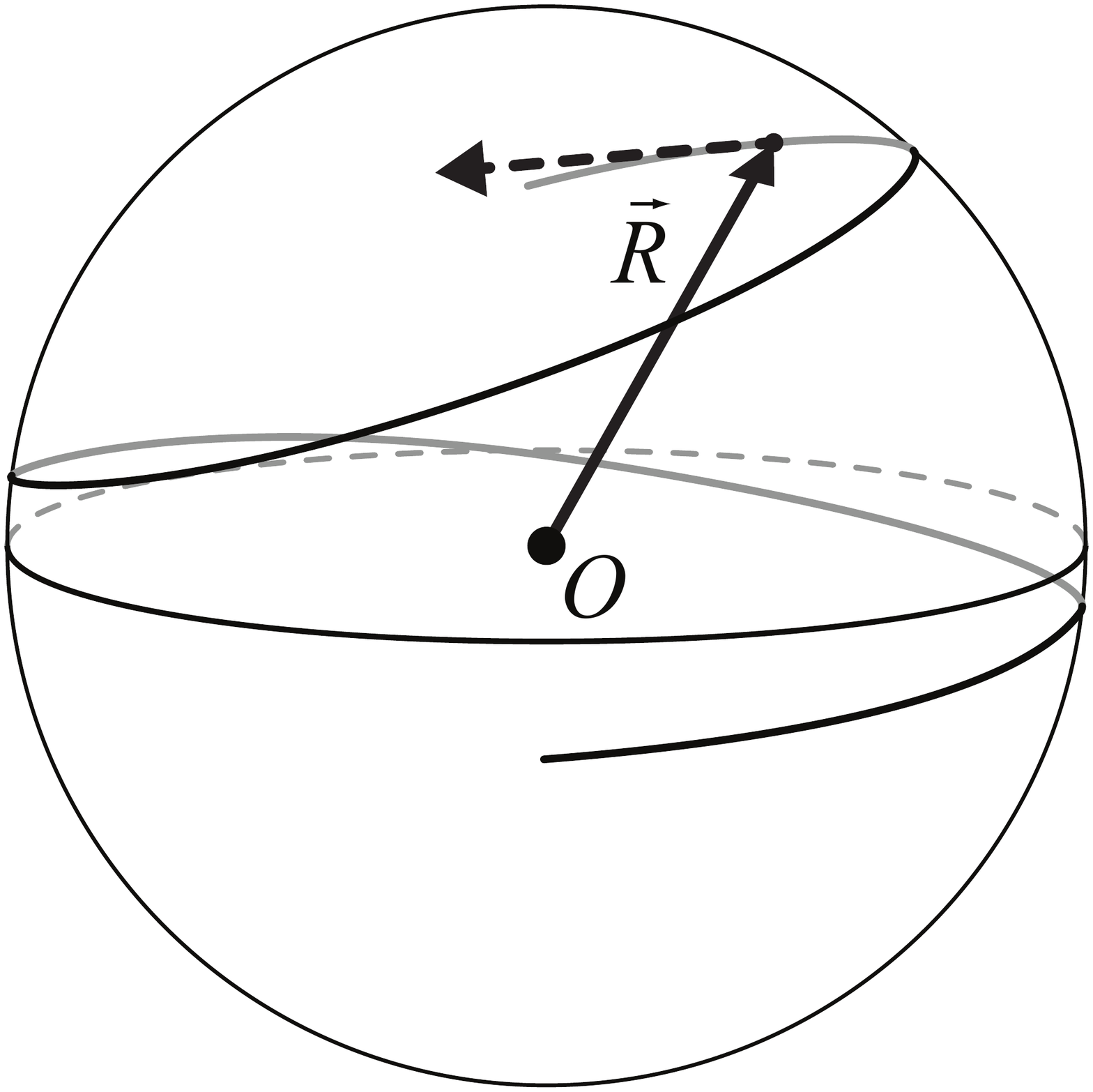}
		\label{fig:fig_constant_length_v01}}
		\caption{Some examples of graphs used in the problems. (a) Level curves of a function of two variables. Several vectors are included to illustrate tangents of the level curves at different arbitrary chosen positions. Also corresponding perpendicular vectors, which are proportional to the gradient, are shown. (b) This graph, that was used in problem P3, had the strongest effect on performance of all the problems in our study. As is discussed in the text, it strongly supports one of the two major possible solution strategies used by the participants involving to depict the dashed vector (not shown for the students), and this solution strategy was very rare in the group not having access to this graph.} 	\label{Fig1}
\end{figure}

The graphs were designed by a lecturer in vector calculus to support students by describing a particular problem-related concept visually. 
In fact all the graphs used in the study could naturally be part of a textbook in vector calculus. Importantly, the students had not seen any of the problems before.
They were interpretational in nature, and should therefore have a substantially positive effect on problem solving. All problems could be solved without having access to the graph. For example, many of the problems can be solved algebraically without using a mental geometric representation. Example graphs can be found in Figure~\ref{fig:fig_Nivakurvor_1_v01}, which shows level curves of a function of two variables and Figure~\ref{fig:fig_constant_length_v01}, which depicts a curve that is restricted to a sphere. Note that the vectors are not labeled, so it was left for the students to identify and potentially use them when testing a particular solution strategy.

Each problem was saved as a grayscale png-image with a resolution of 1680$\times$1050 pixels. This resulted in a total of 16 stimuli images, eight for each group. The problems were presented in a random order.

\subsection*{Apparatus}
The experiment was performed with a Dell laptop (Intel Core i7 CPU 2.67GHz, 2.98 GB RAM) running Windows XP Professional (v. 2002, SP 3). Stimuli were presented with Experiment Center (v. 3.0.128) on a Dell 22 inch computer screen with a resolution of 1680$\times$1050 pixels and a refresh rate of 60 Hz. Eye movements were recorded at 250 Hz with the RED250 eye tracker from SensoMotoric Instruments (Teltow, Germany) running iView X (v. 2.7.13). Data from the left and right eyes were averaged during recording, and therefore only one gaze coordinate represented the data for both eyes at each time instant.

\subsection*{Procedure}
After an introduction to the experiment and after viewing an example problem not included in the actual test, participants were calibrated with a five point calibration followed by a four point validation of the calibration accuracy. Recalibrations were initiated when the operator---watching the eye image in iView X and the stimulus with overlayed gaze position in Experiment Center---judged that it was necessary. 
The average accuracy from all accepted calibrations reported by iView X was 0.5$^{\circ}$ ($\textit{SD}=0.18$) horizontally and 0.55$^{\circ}$ ($\textit{SD}=0.29$) vertically.

Each trial started with a centrally located fixation cross that was presented until the software detected a fixation within a 1$^{\circ}$ square centered on the cross. Then the problem appeared, and the participants were free to inspect the problem for a maximum of 120 seconds. If  they felt that they were ready to provide an answer sooner, they could do so by pressing the spacebar to answer two questions: 
first, participants were asked whether they thought the statement in the problem was \textit{true} or \textit{false} and, second, how certain they were in their answer on a scale from 1 (very unconfident) to 7 (very confident). 
Throughout the eye-tracking experiment, participants were asked to verbalize their thoughts as they solved the problems, according to the methodology described in \citeA[Ch. 3]{holmqvist2011eye} concerning training, instruction, and prompting.

Written consent was given by all participants, who got two movie theater tickets as a compensation for participating. 

\subsection*{Data analysis}
Fixations and saccades were calculated from raw data samples with BeGaze (v. 3.1 Build 152) using default settings.

\begin{figure}
	\centering
\includegraphics[width=0.99\textwidth]{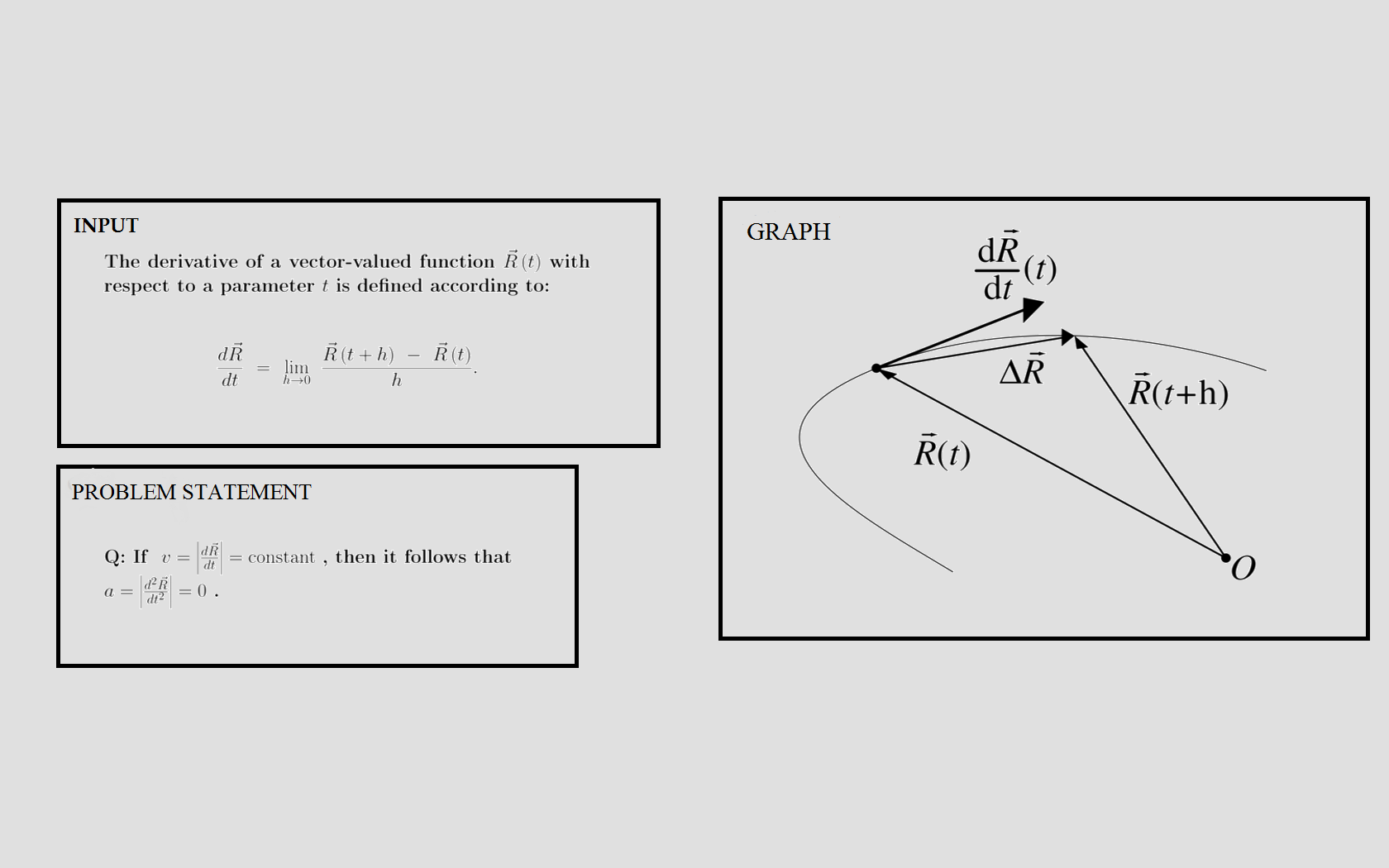}
	\caption{An example of a stimulus (P7) used in the multimedia condition of the experiment, which has three overlayed areas of interest (AOIs): input, problem statement, and graph. } \label{fig:stimuli}
\end{figure}

Eye tracking data were analyzed by means of specific \textit{areas of interest} (AOIs) which we defined for each problem. AOIs are coherent parts of the screen, for which eye tracking parameters were summarized. Figure~\ref{fig:stimuli} depicts a multimedia problem with input, problem statement, and graph, where AOIs are outlined by black rectangles and the name of the AOI is found in the upper left corner of the AOI. AOI names and rectangles were not shown to the participants.
Specifically, we calculated total dwell time (sum of all time spent looking inside an AOI) from raw data samples and transitions between the AOIs from fixation and saccade data.

The proportion of speech was computed from the recorded speech signal, which was sampled at 44 kHz. 
A student was considered to speak when the amplitude ($A$) of the signal exceeded a threshold and when two consecutive speech samples above this threshold were located less than a given number of samples ($n_s$) apart. 
Limits for $A$ and $n_s$ were set to 0.015 (relative intensity) and 440 samples (i.e., 10 ms), respectively. Every part of the speech signal that was not detected as speech by the above definition was considered to be “silence”.

The recorded speech was further analyzed by first transcribing it to text format, and then coding it into `idea units' according to the scheme in Appendix \ref{sec:app} (Table \ref{tab:codingScheme}). 
The main categories in the coding scheme are based on Mayer’s CTML (2005b) and thus refer to the cognitive processes assumed by this theory: searching and selecting of information from input and graph, activating prior knowledge, integrating information from different sources, and the final problem solution. In line with \citeA{van2005uncovering}, who also investigated cognitive processes involved in problem-solving, we included meta-cognitive processes. The actual coding was conducted by two raters for 10\% of all data. Their interrater-reliability was above 70\%, calculated as the number of matching codes with respect to the total number of codes in this 10\% of the data. Since the inter-rater reliability was sufficiently high (i.e., higher than 0.70, \citeNP{van1994think}), one of the raters coded the remaining data.

Data were analyzed with linear mixed effects models using R 2.15.2 \cite{r} and the packages lme4 \cite{lme4} and languageR \cite{languageR}. Participants and problems were modelled as random factors in all analyses.

\section*{Results}
The results are presented in the order of the hypotheses given.

\subsection*{Multimedia effect (H$_1$)}
The participants used on average 90.8 s (\emph{SD} = 28.5) to solve a problem. There was no significant difference in solution time between participants with (\emph{M} = 91.1, \emph{SD} = 28.1) and without  (\emph{M} = 90.5, \emph{SD} = 28.9) graphs (two-sample Kolmogorov-Smirnov test, \emph{p} = 0.76). The difference in solution times were small also when comparing problem where students answer correctly (\emph{M} = 91.0, \emph{SD} = 27.0) and incorrectly (\emph{M} = 90.6, \emph{SD} = 30.2), (two-sample Kolmogorov-Smirnov test, \emph{p} = 0.61).

Participants solving  problems with graphs answered correctly to 56\% 
of the problems compared to 52\% 
for participants without graphs. 
Table~\ref{tab:lmerPerformance} shows the result of a multi-level logistic regression predicting a correct answer based on the presentation condition. 
As can be seen from the table, there is no statistically significant effect of presentation condition on students' abilities to answer the problems correctly.
\begin{table}[ht]
\begin{center}
\caption{Result of the multi-level logistic regression predicting whether the answer was correct or incorrect. Here `withoutgraph' refers to the problems without graphs. The sign of the `Estimate' tells us that the condition with graphs led to a higher proportion of correct answers. However, the effect is not significant since the value of `Pr($>$$|$z$|$)' is above $0.05$.}
\begin{tabular}{lrrrr}
  \hline
 & Estimate & Std. Error & z value & Pr($>$$|$z$|$) \\ 
  \hline
(Intercept) &  0.3019   &  0.4314&   0.700 &   0.484 \\ 
withoutgraph  &-0.1641    & 0.2622 & -0.626  &  0.531 \\ 
   \hline
\end{tabular}
\label{tab:lmerPerformance}
\end{center}
\end{table}

\subsection*{Picture bias (H$_2$)}
\begin{figure}
	\centering
		\includegraphics[width=0.60\textwidth]{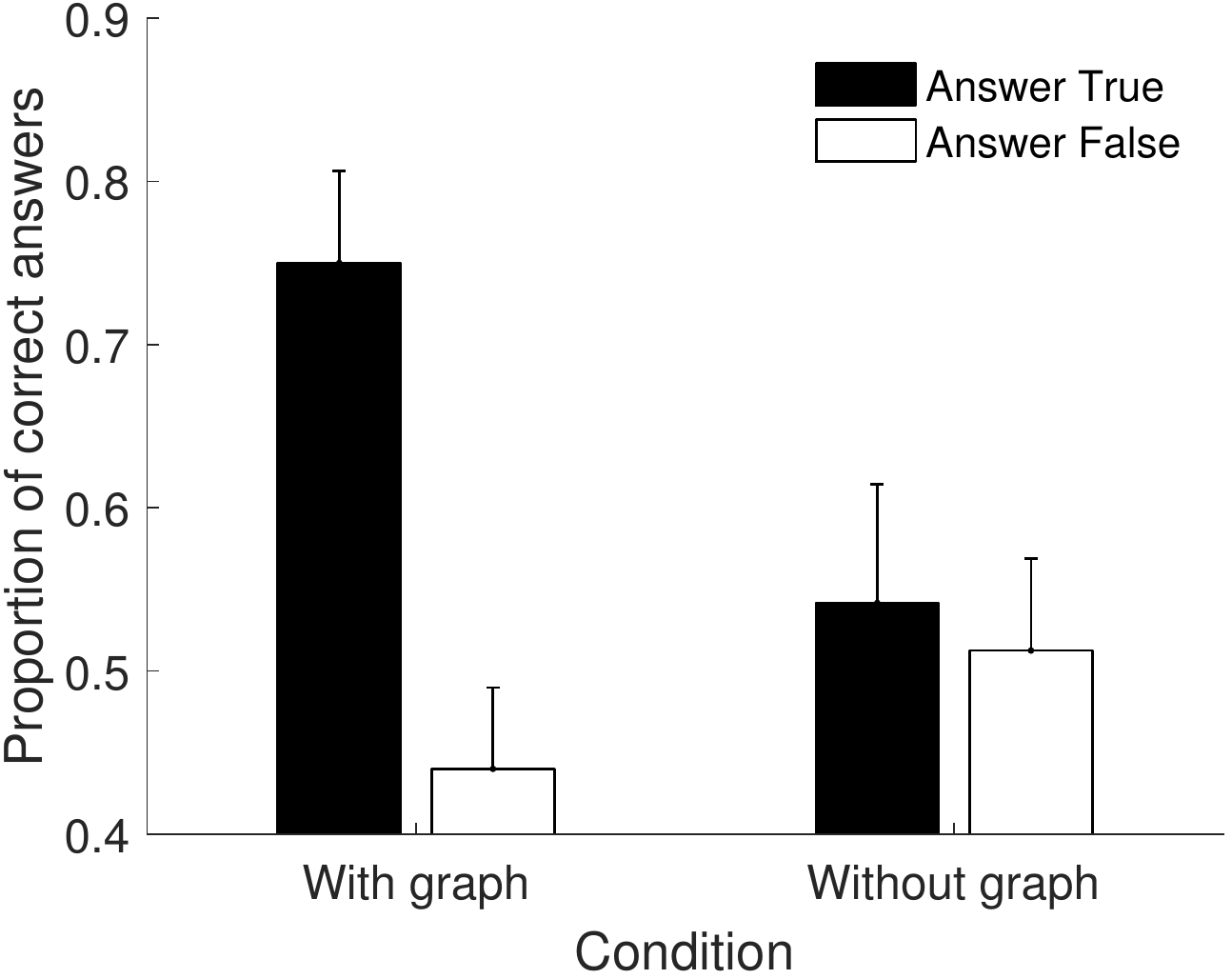}
		\vspace{5mm}
	\caption{Illustration of the interaction between whether the problem statement is true or false in the multimedia (with graph) and control (without graph) conditions. Error bars represent standard errors.}
	\label{fig:interactionPlotMatlab2}
\end{figure}

To test whether there was a confirmation bias when graphs were present, information about  
whether the correct answer is true or false was included in the regression. The output can be seen in Table~\ref{tab:lmerPerformanceCorrectAnswer}.
\begin{table}[ht]
\begin{center}
\caption{Result of the multi-level logistic regression predicting whether the answer was correct or incorrect. Whether the answer is true or false has been included as a factor. Here `*' and `**' indicate significance on the 0.05 and 0.01 levels, respectively.}
\begin{tabular}{lrrrl}
  \hline
 & Estimate & Std. Error & z value & Pr($>$$|$z$|$) \\ 
  \hline
(Intercept)   &              -0.3132 &    0.5121 & -0.612&  0.540  \\ 
withoutgraph    &              0.3537  &   0.3285 &  1.077 & 0.281  \\ 
answerTrue  &            1.7331 &    0.8578 &  2.020 & 0.043* \\
withoutgraph:answerTrue &  -1.4762 &    0.5625 & -2.624 & 0.008**\\
   \hline
\end{tabular}
\label{tab:lmerPerformanceCorrectAnswer}
\end{center}
\end{table}
The analysis reveals that participants were more likely to answer the problem statement correctly if it is true and, interestingly, there is a significant interaction between presentation condition and whether the answer is true or false. As illustrated in Figure \ref{fig:interactionPlotMatlab2}, it appears as if the students were more likely to answer correctly when the answer was true and a graph was present, compared to when the answer was false. On the contrary, whether the answer was true or false had no influence when the graph was not present. A post-hoc multiple comparison\footnote{With the R-package \texttt{multcomp} and the function \texttt{glht} using Tukey contrasts to compare the means.} revealed only one marginally significant difference, which occurred between the two conditions when the answer is true ($p=0.056$). Additional support for a picture bias is provided in Table \ref{tab:lmerPerformanceAnswerTrue}, which shows that presentation condition is a significant predictor for providing a confirmatory answer.
\begin{table}[ht]
\begin{center}
\caption{Result of the multi-level logistic regression predicting the answer (confirm/reject) based on presentation condition. The sign of the estimate for 'withoutGraph' tells us that more confirmatory answers were given in the multimedia condition.}
\begin{tabular}{lrrrl}
  \hline
 & Estimate & Std. Error & z value & Pr($>$$|$z$|$) \\ 
  \hline
(Intercept)   &              0.7224  &   0.4358  & 1.658  & 0.0974  \\ 
withoutgraph    &              -0.6366   &  0.2859 & -2.226 &  0.0260*  \\ 
   \hline
\end{tabular}
\label{tab:lmerPerformanceAnswerTrue}
\end{center}
\end{table}

Since the nature of the answer (true or false) turned out to significantly predict the proportion of correct answers, this predictor was included in all further statistical models.

\subsection*{Information processing}
\subsubsection*{Search and selection (H$_{3a}$)}
The overall small effect graphs had on comprehension raises the question of how the students utilize the additional graphical information. 
Given the similar performance results, it is tempting to believe that they did not spend much time on the graphs but, as in the non-illustrated condition, inspected only the text and the equations in the input and problem statement areas. 
At the same time, the interaction between whether the answer was true or false and the presentation condition (with or without graph) suggests that the graph influenced the students' problem solving processes.
\begin{figure}
	\centering
		\includegraphics[width=0.80\textwidth]{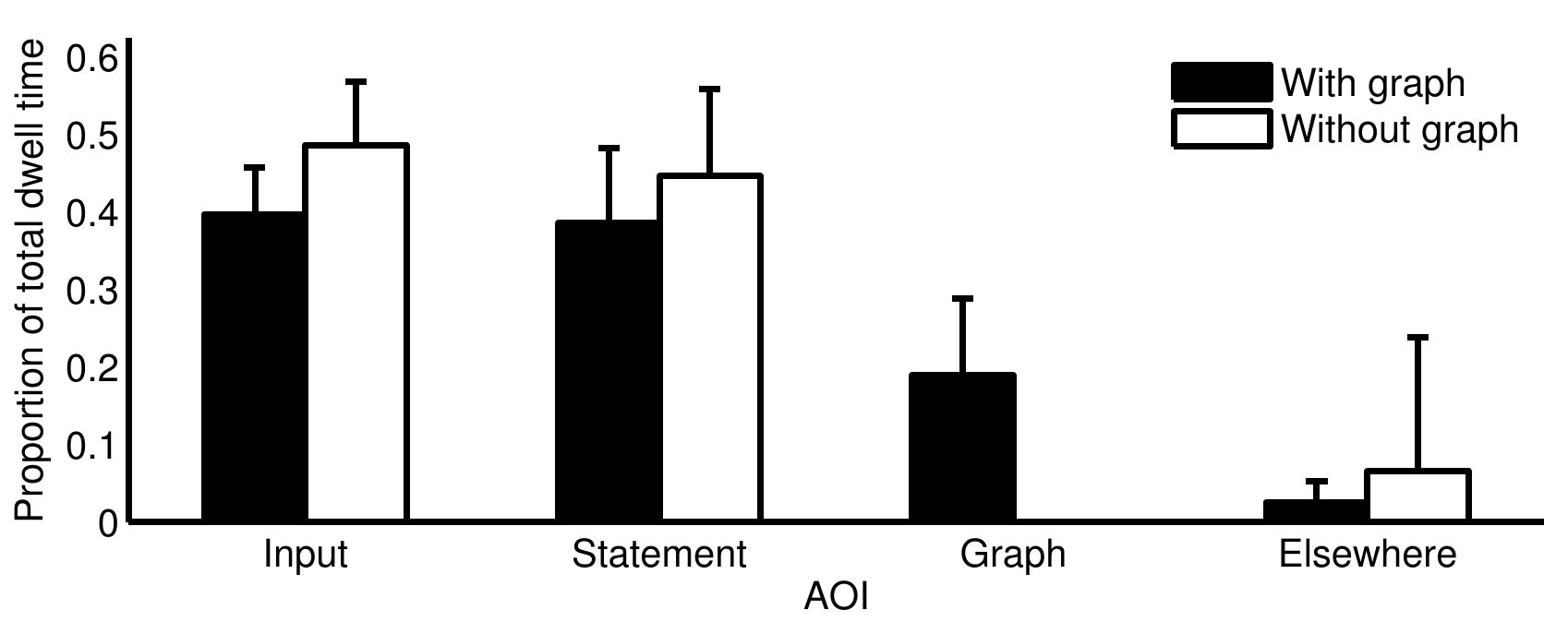}
	\caption{Proportion of total dwell time with and without graphs. Error bars represent standard errors.}
	\label{fig:Prop_totDwellTime}
\end{figure}
Overall, the students spend a fairly large proportion of the total time viewing the graphs (19.0$\pm$10.2\%). As can be seen from Figure \ref{fig:Prop_totDwellTime}, it appears as if the graph is inspected at the expense of the input and the problem statement, in such a way that equal amounts of time is taken from each of these regions. The proportion of total dwell time on both the input and problem statement was significantly shorter when the graph was present, according to a two sample \emph{t}-test ($p<0.001$).
Moreover, a similar test between the quotients of `input' and `problem statement' for problems with and without graphs did not come out significant (\emph{p} > 0.05) for any of the problems. 

\begin{figure}
	\centering
		\includegraphics[width=0.60\textwidth]{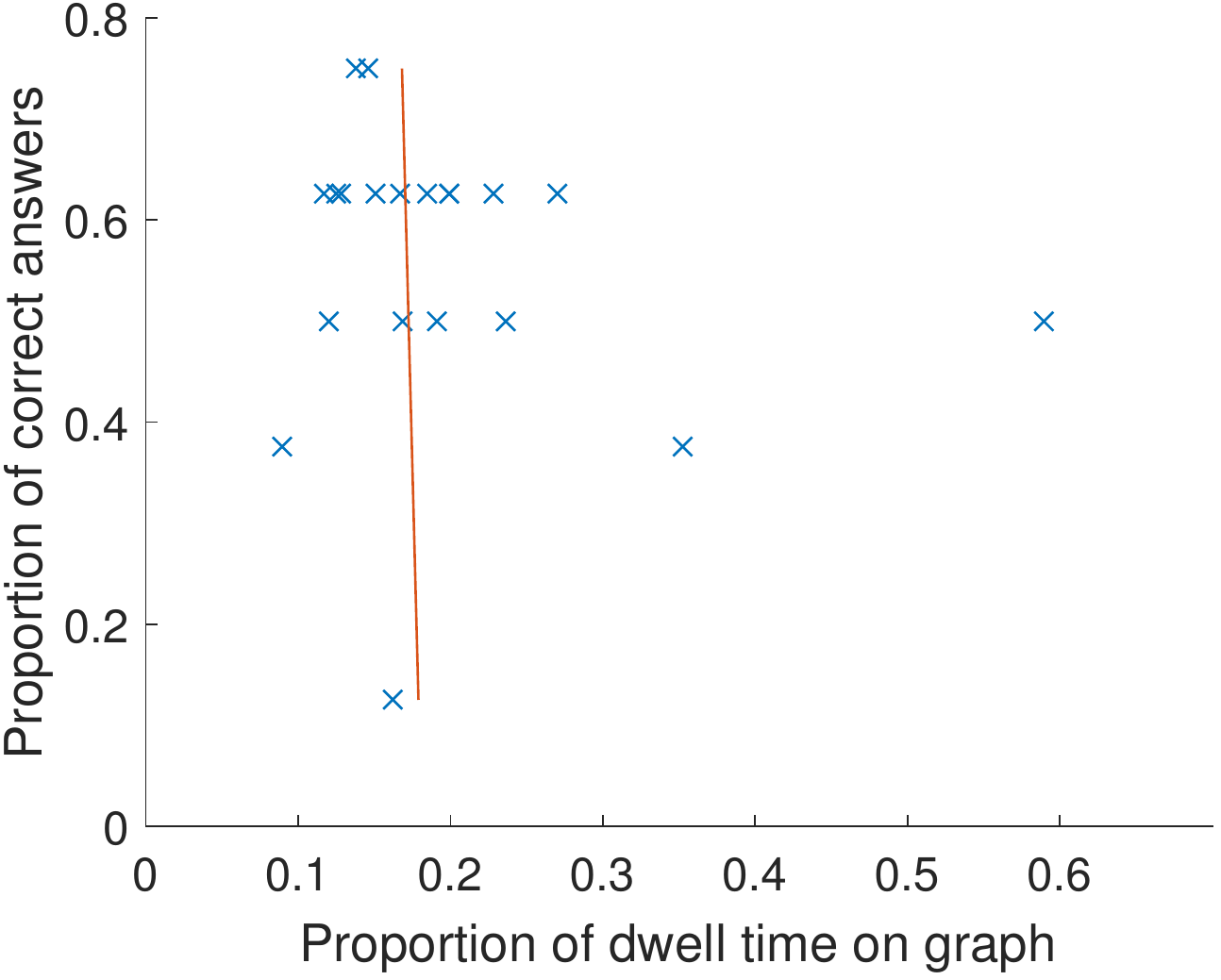}
	\caption{The relationship between total dwell time on the graph and the proportion of correct answers. Each $\times$ corresponds to one of the 20 participant in the multimedia condition, and the line represents a linear fit of the data.}
	\label{fig:performanceVsDwellTimeOnGraph}
\end{figure}

Given that we know that a significant portion of time is spent visually inspecting the graph, does a longer inspection time also lead to better performance? On average, there were small differences in total dwell time on the graph when answering correctly ($M=19.9, SD=11.0$~\%) compared to incorrect answers ($M=18.6, SD=10.8$~\%) and, as seen in Figure~\ref{fig:performanceVsDwellTimeOnGraph}, there was no relationship between whether participants answered correctly and how much time they spent looking at the graph. 
This is confirmed statistically by the results reported in Table~\ref{tab:lmerPerformanceVsDwellTimeOnGraph}.
\begin{table}[ht]
\begin{center}
\caption{Result of the multi-level logistic regression predicting whether the answer was correct or incorrect. The proportion of dwell time on the graph is included as a factor. Note that the proportion of dwell time is logit-transformed before being used in the model.}
\begin{tabular}{lrrrl}
  \hline
 & Estimate & Std. Error & z value & Pr($>$$|$z$|$) \\ 
  \hline
(Intercept)             &  -0.5382 &    0.6945 & -0.775 & 0.43840   \\ 
logit(propDwellTime)    &  -0.1280  &   0.2754 & -0.465 & 0.64195   \\ 
answerTrue              & 3.6997    & 1.2825  & 2.885&  0.00392 ** \\
logit(propDwellTime):answerTrue &   1.1224   &  0.4765  & 2.355 & 0.01850 * \\
   \hline
\end{tabular}
\label{tab:lmerPerformanceVsDwellTimeOnGraph}
\end{center}
\end{table}

\begin{figure}
	\centering
		\includegraphics[width=0.60\textwidth]{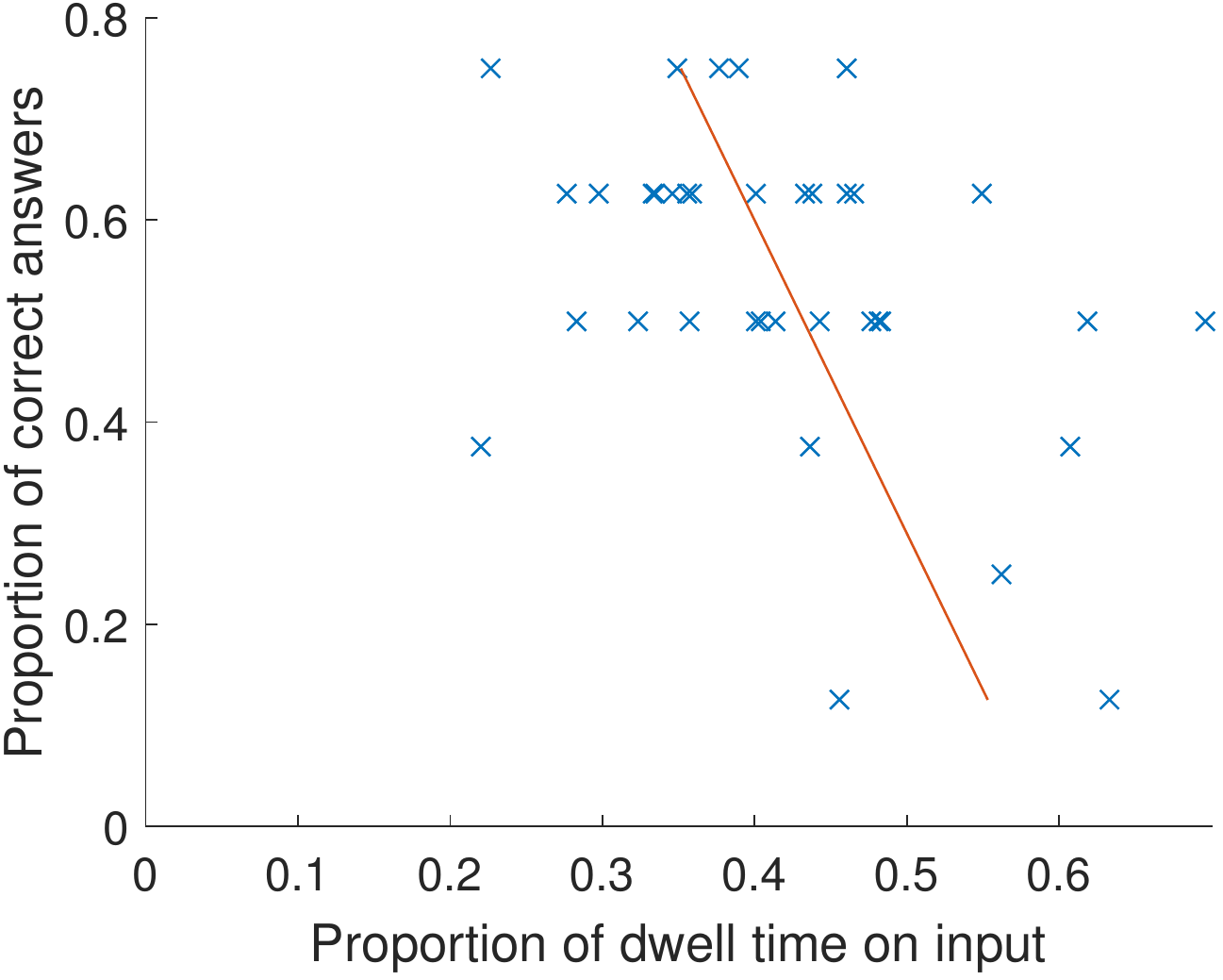}
	\caption{The relationship between total dwell time on the input and the proportion of correct answers. Each $\times$ corresponds to one of the 36 participants, and the line represents a linear fit of the data.}
	\label{fig:performanceVsDwellTimeOnInput}
\end{figure}

Previous research has shown that a good problem solving strategy is to read the problem formulation  carefully, before moving on to other parts of the problem~\cite{andra2009students}. However, Figure \ref{fig:performanceVsDwellTimeOnInput} shows that performance is inversely proportional to the proportion of dwell time on the input area in a problem. 
Students who answered correctly looked at the input 42.6\% ($SD=11.1$) of the time whereas those who answered incorrectly spent 47.0\% ($SD=12.1$) of the time inspecting the input.
A smaller proportion of dwell time on the input significantly predicts an increase in performance (\emph{cf.} Table \ref{tab:lmerPerformanceVsDwellTimeOnInput}).
\begin{table}[ht]
\begin{center}
\caption{Result of the multi-level logistic regression predicting whether the answer was correct or incorrect. The proportion of dwell time on the input is included as a factor. Note that the proportion of dwell time is logit-transformed before being used in the model.}
\begin{tabular}{lrrrl}
  \hline
 & Estimate & Std. Error & z value & Pr($>$$|$z$|$) \\ 
  \hline
(Intercept)             &  -0.4317 &     0.5543 & -0.779 & 0.4361   \\ 
logit(propDwellTime)    &  -0.5694  &   0.2618 & -2.175 & 0.0296 *   \\ 
answerTrue              & 1.2926   &  0.9021 &  1.433 &  0.1519 \\
logit(propDwellTime):answerTrue &    -1.0773  &   0.5259 & -2.049 &  0.0405 * \\
   \hline
\end{tabular}
\label{tab:lmerPerformanceVsDwellTimeOnInput}
\end{center}
\end{table}

\begin{figure}
	\centering
		\includegraphics[width=0.60\textwidth]{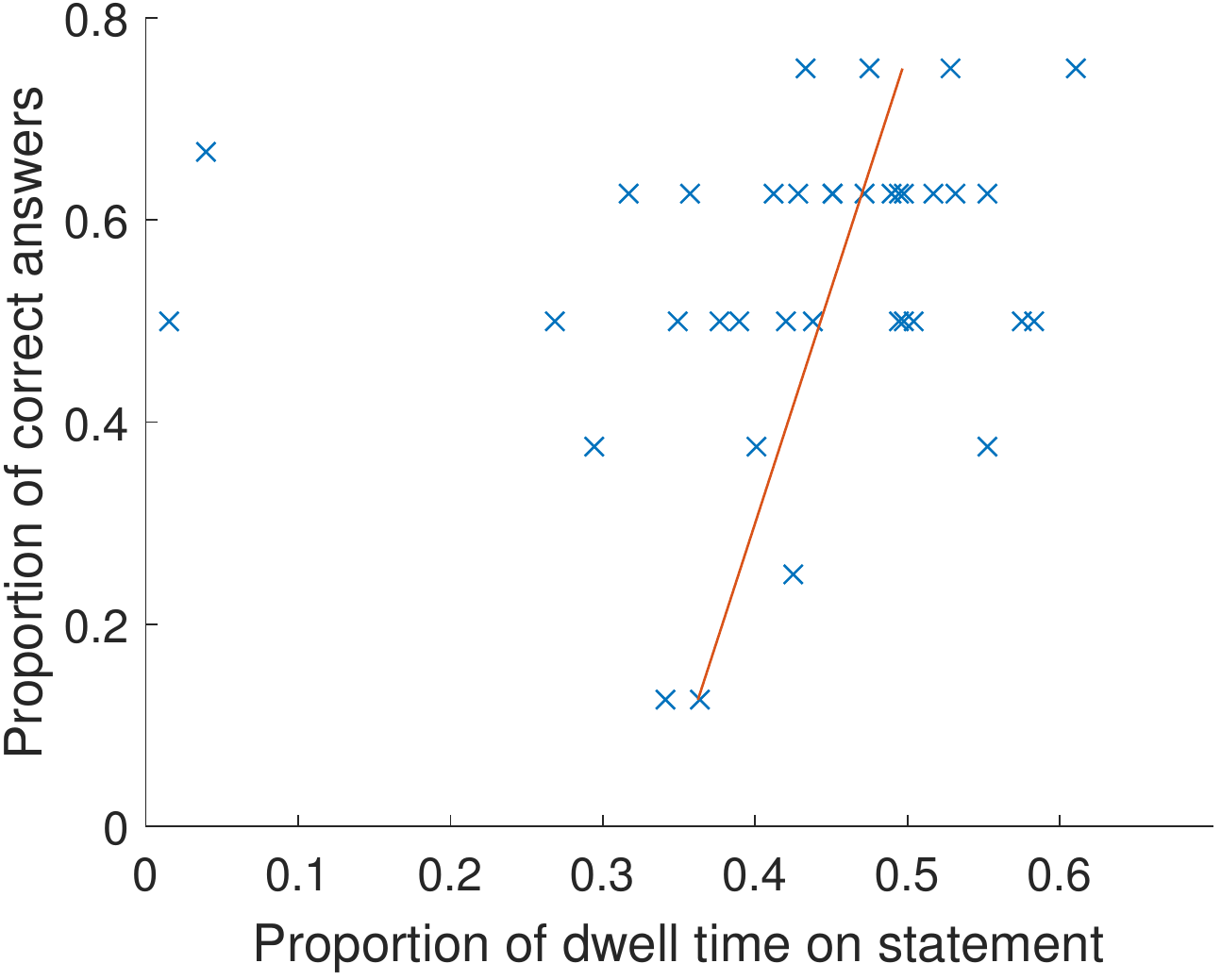}
	\caption{The relationship between total dwell time on the problem statement and the proportion of correct answers. Each $\times$ corresponds to one of the 36 participants from both conditions, and the line represents a linear fit of the data.}
	\label{fig:performanceVsDwellTimeOnQuestions}
\end{figure}

As shown in Figure~\ref{fig:performanceVsDwellTimeOnQuestions}, the students dwelled proportionally longer at the problem statement when giving a correct answer ($M=42.6, SD=12.8$~\%) compared to an incorrect answer ($M=38.3, SD=11.3$~\%), and the total dwell time on the statement was a significant predictor for a correct answer (\emph{cf.} Table~\ref{tab:lmerPerformanceVsDwellTimeOnQuestion}).
\begin{table}[ht]
\begin{center}
\caption{Result of the multi-level logistic regression predicting whether the answer was correct or incorrect. The proportion of dwell time on the problem statement is included as a factor. Note that the proportion of dwell time is logit-transformed before being used in the model.}
\begin{tabular}{lrrrl}
  \hline
 & Estimate & Std. Error & z value & Pr($>$$|$z$|$) \\ 
  \hline
(Intercept)             &  -0.01989  &  0.49089 & -0.040&  0.96768   \\ 
logit(propDwellTime)    &  0.55821   & 0.21766  & 2.565  &0.01033 *   \\ 
answerTrue              &   0.56485  &  0.82891 &  0.681 & 0.49560 \\
logit(propDwellTime):answerTrue &  -1.09686 &   0.41480 & -2.644 & 0.00818 **\\
   \hline
\end{tabular}
\label{tab:lmerPerformanceVsDwellTimeOnQuestion}
\end{center}
\end{table}

\subsubsection*{Integration (H$_{3b}$)}
\begin{figure}
	\centering
		\includegraphics[width=0.60\textwidth]{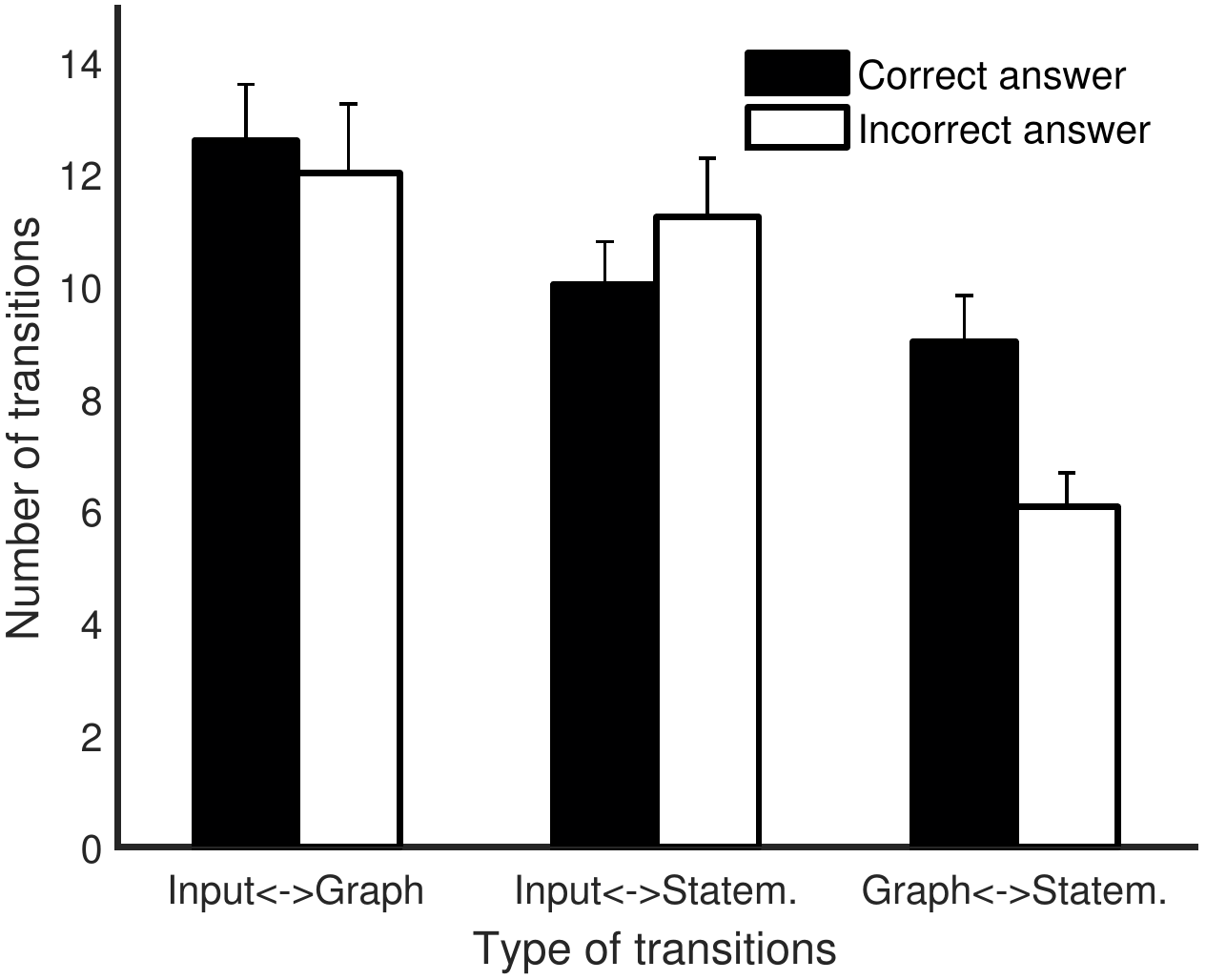}
	\caption{Number of transitions between the different AOIs: input, graph, and problem statement. Transitions in both direction are included. Error bars represent standard errors.}
	\label{fig:performanceVsnTransitions}
\end{figure}

It could be that a long dwell time on the graph by itself does not help students' problem solving, but rather how they integrate the graph with other parts of the problem, i.e., the regions labeled as \textit{input} and \textit{problem statement} before.
Figure~\ref{fig:performanceVsnTransitions} illustrates how performance is related to the number of transitions between different areas in the problem. 
As shown in Table~\ref{tab:lmerPerformanceVsnTransitions}, there is a marginally significant effect ($p=0.08$) that the number of transitions between the graph and the problem statement were higher for students that answered a problem correctly. No significant differences were found for the other transitions in Figure~\ref{fig:performanceVsnTransitions}. 

\begin{table}[ht]
\begin{center}
\caption{Result of the multi-level logistic regression predicting whether the answer was correct or incorrect. The number of transitions between graph and problem statement is included as a factor. Here `.' indicates that an effect is marginally significant.}
\begin{tabular}{lrrrl}
  \hline
 & Estimate & Std. Error & z value & Pr($>$$|$z$|$) \\ 
  \hline
(Intercept)             &  -0.84588  &  0.51767 & -1.634 &  0.1023   \\ 
nTransitions     &  0.06811   & 0.03928 &  1.734  & 0.0829 .  \\ 
answerTrue              & 1.48012  &  0.85270 &  1.736 &  0.0826 .\\
withoutgraph:answerTrue &  0.01943   & 0.07887 &  0.246  & 0.8054\\
   \hline
\end{tabular}
\label{tab:lmerPerformanceVsnTransitions}
\end{center}
\end{table}

\subsubsection*{Mental effort measured with verbal data  (H$_{3c}$)}
\begin{figure}
	\centering
\includegraphics[width=0.7\textwidth]{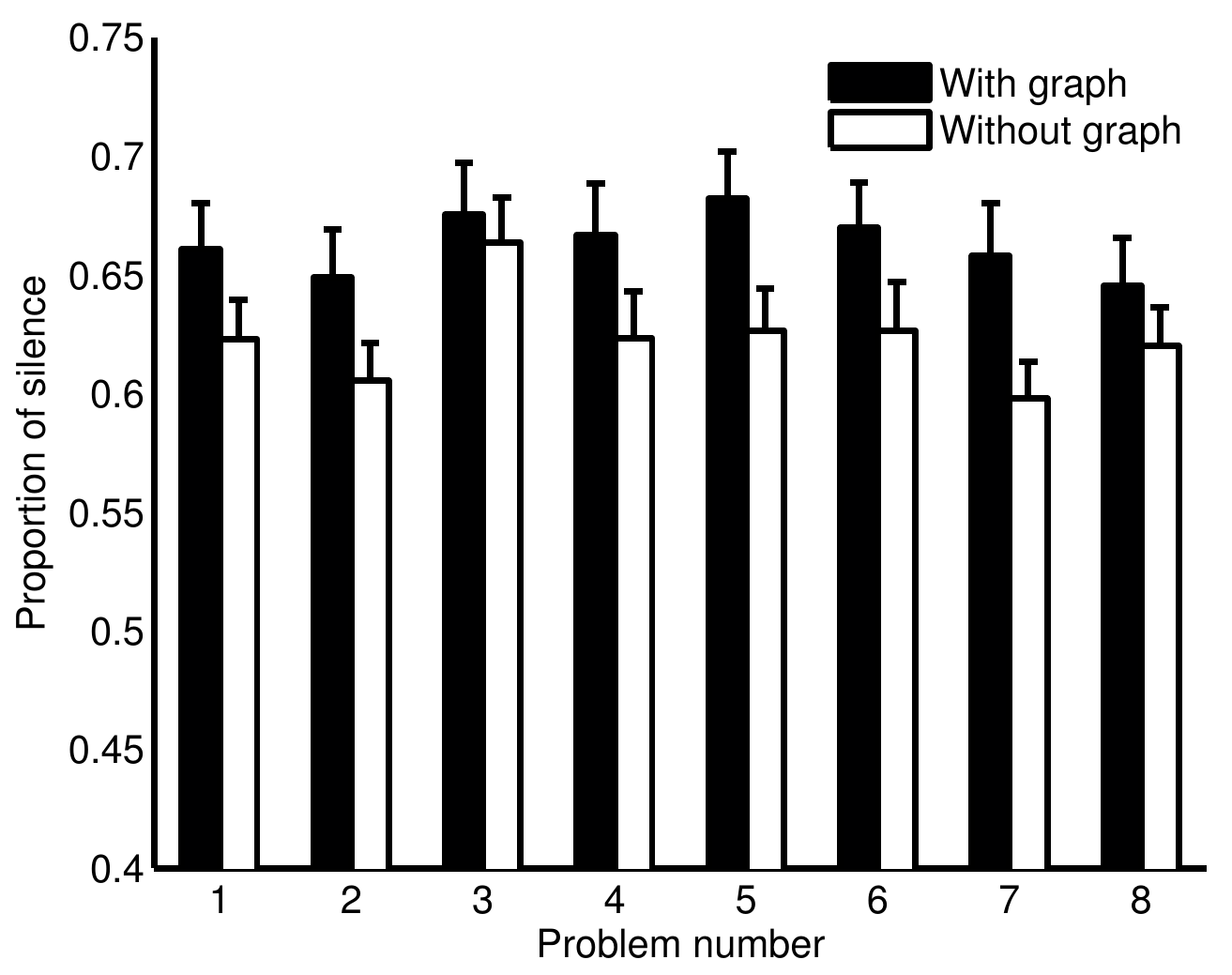}		
\caption{Proportion of silence used as a measure of mental effort. Error bars represent standard errors. }		
	\label{fig:mentalEffort}
\end{figure}

To estimate whether the illustrated problems required more mental effort, the proportion of silence was calculated from the verbal data. 
Figure~\ref{fig:mentalEffort} shows that participants consistently speak  less when the problem includes a graph; the proportion of silence increases from 62.3\% ($SD=7.0$) to 66.3\% ($SD=9.1$) for participants in the multimedia condition. 
As shown in Table \ref{tab:lmerPerformanceVsMentalEffort}, there is a marginally significant effect of presentation condition on the proportion students speak (\emph{p} $= 0.07$). 
\begin{table}[ht]
\begin{center}
\caption{Result of the multi-level logistic regression predicting the proportion of silence.}
\begin{tabular}{lrrrr}
  \hline
 & Estimate & Std. Error & t value &  Pr($>$$|$t$|$)\\ 
  \hline
(Intercept)             &   0.662  			&  0.016 		& 39.31  & <2e-16\\ 
withoutgraph     				&  -0.044   		& 0.023 		& -1.86  & 0.071 \\ 
answerTrue              & 	0.002  			&  0.012 		&  0.18  & 0.863 \\
withoutgraph:answerTrue &  	0.012   		&  0.012 		&  0.246 & 0.319 \\
   \hline
\end{tabular}
\label{tab:lmerPerformanceVsMentalEffort}

\end{center}
\end{table}

\subsection*{Results verbal data - two contrasting cases (RQ$_1$)}
\begin{figure}
	\centering
\subfigure[Problem 3. Input: A vector-valued function $\overset \rightarrow R(t)$ has the property: $|\overset \rightarrow R(t)|^2 = \overset \rightarrow R(t) \cdot \overset \rightarrow R(t) = $ constant. Problem statement (Q): Then it follows that $\overset \rightarrow R(t) \bot \frac{d\overset \rightarrow R(t)}{dt}$.]{\includegraphics[width=0.99\textwidth]{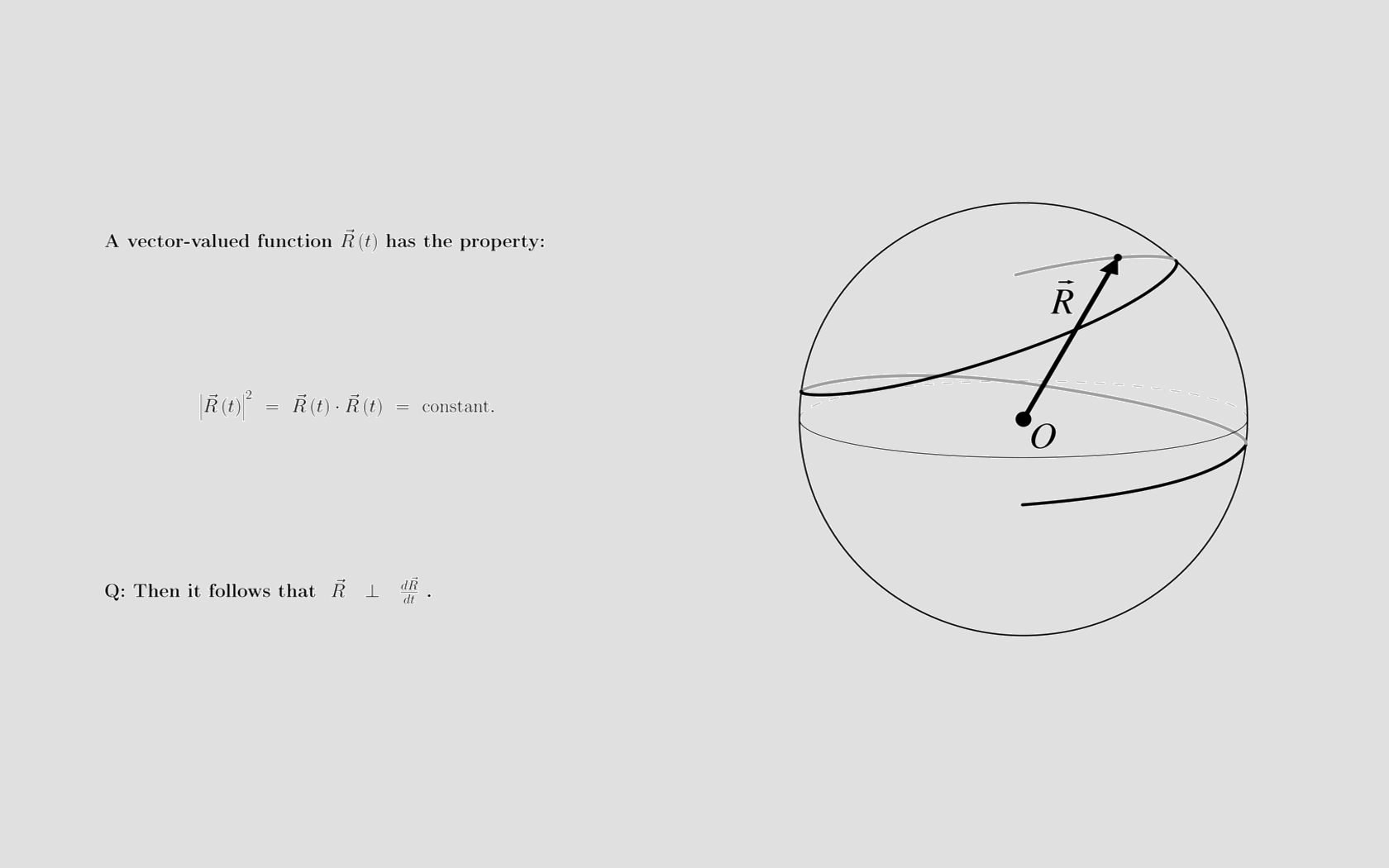}		}
\subfigure[Problem 4. Input: Let $\overset \rightarrow v$ be a smooth vector-field defined in $\mathbb{R}^3$. The compact volume $V$ has a boundary $\partial V$, composed of one or several smooth surfaces, and oriented with an outbound normal $\overset \rightarrow {n}$. Then: $\iint_{\partial V} \overset \rightarrow v \cdot \overset \rightarrow n dS = \iiint_V \nabla \cdot \overset \rightarrow v dV$. Problem statement (Q): Then the so called Gauss formula have the interpretation that the production of a scalar quantity integrated over the volume $V$ is equal to the surface integral of the corresponding gradient.]{\includegraphics[width=0.99\textwidth]{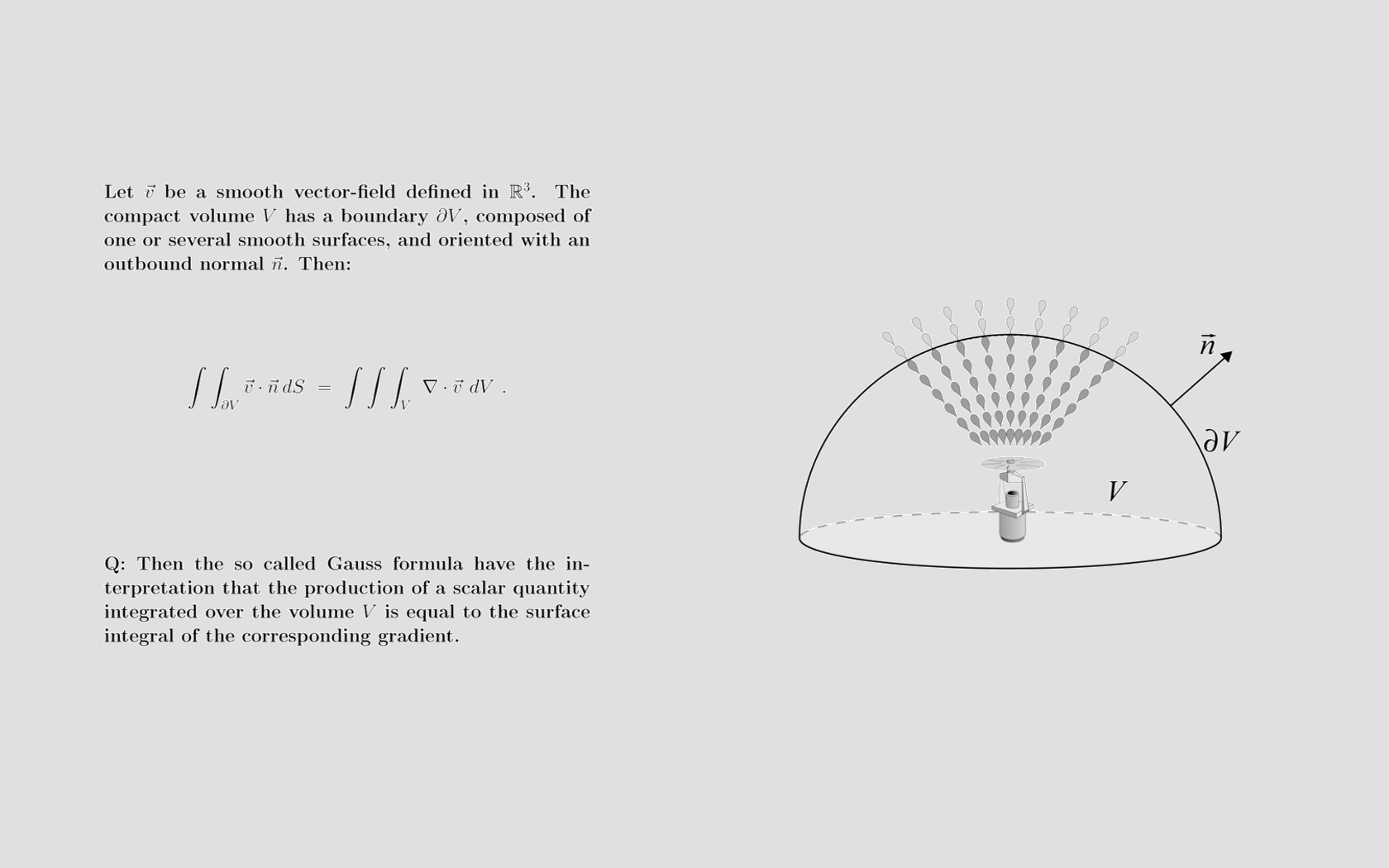}		}
\caption{Two contrasting cases. (a) P3, including the graph that improved the performance the most, and (b) P4, including the graph that helped the least. For improved readability, the text in the stimuli has been reproduced in the figure captions.}
	\label{fig:twocases}
\end{figure}

In this section we compare the two most extreme cases from our experiment with respect to verbal data: problem P3 for which the presence of a graph improved the results the most, and problem P4 for which the results for the group having access to the graph was the worst. These problems are shown in Figure~\ref{fig:twocases}. In P3, the graph increased the proportion of correct answers from 0.25 ($SD=0.44$) to 0.70 ($SD=0.47$), whereas the graph in P4 decreased the proportion of correct answers from 0.38 ($SD=0.50$) to 0.10 ($SD=0.31$).
Here we report analyses of verbal data based on the coding schema described in Appendix \ref{sec:app} (Table \ref{tab:codingScheme}).
\begin{sidewaystable}
	\begin{center}
	\parbox{7in}{\caption{Results of the verbal coding for problems 3 and 4. Each number represents the number of codes for all participants divided by the number of participants. `Diff' represents the difference WithGraph - WithoutGraph for Problem 3, Problem 4, and Problem 3 + Problem 4. The numbers in the last column represent differences between the two problems in the multimedia condition (with a graph). For definitions and explanations of the codes, cf. Appendix \protect{\ref{sec:app}}. 
	Color-coded codes represent the largest differences between the multimedia and control condition. Blue color indicates differences unrelated to graphs whereas red color indicates differences related to graphs. To make the table more readable and compact, WithGraph is denoted 'Graph' and WithoutGraph is denoted 'noGraph'.}\label{tab:verbalCodes}
}
	\footnotesize
	\vspace{-9cm}
\begin{tabular}{l|ccc|ccc|ccc|c}
\hline
Code & \multicolumn{3}{ c }{P3} &\multicolumn{3}{ c }{P4}&\multicolumn{3}{ c }{Total (P3+P4)} & P3$_{\text{Graph}}$-P4$_{\text{noGraph}}$\\ \hline
 &Graph &noGraph & Diff &Graph &noGraph& Diff&Graph &noGraph & Diff & \\
\hline
S1T&3.00&3.38&-0.38&5.80&5.56&0.24&8.80&8.94&-0.14&\textcolor{blue}{-2.80}\\
S1G&0.90&0.00&0.90&0.35&0.00&0.35&1.25&0.00&\textcolor{red}{1.25}&\textcolor{red}{0.55}\\
S1S&1.50&1.31&0.19&5.10&4.62&0.47&6.60&5.94&\textcolor{blue}{0.66}&\textcolor{blue}{-3.60}\\
MMVT&0.45&0.69&-0.24&0.10&0.06&0.04&0.55&0.75&-0.20&\textcolor{blue}{0.35}\\
MMVS&0.05&0.00&0.05&0.00&0.00&0.00&0.05&0.00&0.05&0.05\\
MMG&0.60&0.00&0.60&0.05&0.00&0.05&0.65&0.00&\textcolor{red}{0.65}&\textcolor{red}{0.55}\\
PK&3.30&4.06&-0.76&3.05&3.38&-0.33&6.35&7.44&\textcolor{blue}{-1.09}&\textcolor{blue}{0.25}\\
ITG&0.40&0.00&0.40&0.50&0.00&0.50&0.90&0.00&\textcolor{red}{0.90}&-0.10\\
ITS&0.20&0.56&-0.36&0.55&0.69&-0.14&0.75&1.25&\textcolor{blue}{-0.50}&\textcolor{blue}{-0.35}\\
IGS&0.50&0.00&0.50&0.00&0.00&0.00&0.50&0.00&\textcolor{red}{0.50}&\textcolor{red}{0.50}\\
EPS&0.95&0.69&0.26&0.90&0.88&0.03&1.85&1.56&\textcolor{blue}{0.29}&0.05\\
SS&0.35&0.12&0.22&0.30&0.38&-0.08&0.65&0.50&0.15&0.05\\
TS&0.40&0.19&0.21&0.55&0.19&0.36&0.95&0.38&\textcolor{blue}{0.58}&-0.15\\
GS&0.05&0.00&0.05&0.00&0.00&0.00&0.05&0.00&0.05&0.05\\
MCT+&0.20&0.25&-0.05&0.35&0.25&0.10&0.55&0.50&0.05&-0.15\\
MCT-&0.00&0.12&-0.12&0.00&0.06&-0.06&0.00&0.19&-0.19&0.00\\
MSO+&0.40&0.06&0.34&0.10&0.12&-0.02&0.50&0.19&\textcolor{blue}{0.31}&\textcolor{blue}{0.30}\\
MSO-&1.75&1.69&0.06&1.15&1.25&-0.10&2.90&2.94&-0.04&\textcolor{blue}{0.60}\\
GO&0.25&0.19&0.06&0.30&0.31&-0.01&0.55&0.50&0.05&-0.05\\
OT&0.35&0.19&0.16&0.55&0.69&-0.14&0.90&0.88&0.03&-0.20\\
\end{tabular}
\end{center}
\end{sidewaystable}
Table~\ref{tab:verbalCodes} shows the normalized frequency of each code in relation to the two contrasting problems.
In addition to the verbal analysis, we report how confident students were in their answers. 

\subsubsection*{Effect of graph presence}

To estimate the effect of the graph, we have calculated the difference between the codes in the multimedia and the control condition across both problems. Then, we picked the ten largest differences between these. 
Four of these differences were related to the graph. In that, we found that when a graph was present participants selected more information from the graph (1.25), they integrated more information from the graph with the statement (0.50) and with the input (0.90). 
Furthermore, they built more mental models based on the graph (0.65). 
Hence, the participants made more active use of the graphs. 
However, the presence of a graph did not only influence its use, but also the use of the other problem elements. 
In that, we found positive and negative influences of the graph. 
On the positive side for the performance in the multimedia condition, participants selected more information from the statement when a graph was present (0.66).
Since we found that a proportionally longer dwell time on the statement was related with higher performance, more information selection from the statement can be seen as a positive effect of the graph. 
Moreover, the participants evaluated to a higher extent whether or not the statement (0.29) and the input were correct (0.58). 
At the same time, they used less prior knowledge (-1.09) and they integrated information with the input and the statement less frequently (-0.50), which in turn probably is a negative effect of the graph. Furthermore, participants evaluated their own knowledge more positively (0.31), which may also be problematic. 
In summary, adding a graph seems to have both positive and negative effects on the processes underlying problem solving.

\subsubsection*{Effect of helpfulness of the graph}
To further investigate the effects of a graph, we compared the use of graphs for two contrasting cases: when the graph was most helpful and when it was most harmful. Therefore we calculated the difference between the two problems in the multimedia condition. Again, we chose the ten biggest differences. Three of these differences were directly 
related to the use of graphs. 
We found that when the graph was helpful, it was selected to a higher extent (0.55), more integrated with prior knowledge (0.5),
and participants built more mental models from it (0.55) compared to when the graph was harmful. Hence, participants made a more active use of the graph, when it was helpful. 
Moreover, we found also impacts of the graph on processing the other information sources: when the graph was helpful, participants selected less information from the input (-2.8) and from the statement (-3.6), integrated less information from the statement and the input (-0.35), but built more mental models based on the input (0.35). Thus, when the graph was helpful, participants extracted less information from other sources, but still used these more actively. 
Moreover, when the graph was helpful, participants evaluated their own knowledge more---both in positive (0.3) and negative (0.6) terms. 

Furthermore, we conducted a more qualitative analysis of these two contrasting cases.

\textit{Problem 3: When the graph was most helpful}. By quantifying how often the keyword \textit{sphere} (or \textit{circle}) occurs in the verbal data, it seems as if the graph [see Figure~\ref{fig:fig_constant_length_v01}] directly supported the most common way of solving this problem. 
That is to mentally picture the dashed arrow in Figure~\ref{fig:fig_constant_length_v01} (or the opposite oriented counterpart), resulting from the vector $ \vec{R} \left( t \right) $ moving along the trajectory, and to finally recognize that the dashed vector is tangential to the sphere and hence orthogonal to its radius. In the multimedia condition 50\% of the participants used the keyword while reasoning about the problem, and 90\% of these gave the correct answer `true'.  In the control group only 6\% uttered the keyword. 

The confidence scores for problem P3 support the view that a majority of the participants who answers true actually has solved the problem correctly; among the students having access to the graph, there was a higher confidence ($M = 5.3$) for students who answered true, which is the correct answer, than for those who answered false ($M = 4.3$). Similarly, for the group not having access to the graph, the confidence ($M = 4.6$) was also higher for those who answered the problem correctly than those who did not ($M = 4.0$). In summary, it seems that participants in the multimedia condition to a large extent actively used the graph to solve this problem, and were also confident about their solutions.

\textit{Problem 4: When the graph was most harmful}. The graph in P4, related to Gauss formula, illustrates that material being created within a volume is equal to the flow of material through the boundary of that volume. Hence, this interpretation of the Gauss formula is expected to be clearer for the group having access to the graph. In the verbal analysis we found that 25\% of the participants in the multimedia condition commented on this interpretation, while in the control group this number decreased to 19\%. More interestingly, 
we found that 60\% in the multimedia condition said (something similar to) \textit{this must be correct}, while in the control condition such statements were uttered only by 19\% of the participants. 

Turning to the confidence scores we found that, for students in the multimedia condition, there was a higher confidence ($M=4.4$) among students who answered true, which is the wrong answer, than for those who correctly answered false ($M=3.0$). On the contrary, in the control condition, the confidence ($M=4.6$) was higher for those who answered the problem correctly (i.e., false), than those who did not ($M=4.0$).

Taken together, participants seems to be more likely---and confident---to confirm a statement, when a difficult problem is accompanied by a graph.

\section*{Discussion}
In this study we investigated the multimedia effect in problem solving at the university level with examples taken from the field of vector calculus. We found no support for an overall multimedia effect (H$_1$). Instead, graphs had a beneficial effect on performance only when problem statements were to be confirmed (instead of rejected), which is referred to as the picture bias effect (H$_2$). With respect to H$_3$, analyses of eye movement data showed that the graphs attracted students' visual attention at the expense of fewer looks toward other parts of the problem. 
Moreover, spending a proportionally long time inspecting the problem statement as well as frequently moving the eyes between the graph and the problem statement correlated with a higher performance.

Finally, analyses of verbal data provided further insights into why graphs can be both helpful and harmful. 
It was hypothesized (H$_3$) that the students would actively use the graph in terms of utterances relating to the graph (H$_{3a}$) and integration between the graph and other information sources (H$_{3b}$).
Results showed that when a graph was present participants indeed made active use of it, both in terms of selection and integration, and even more so when the graph was helpful. Interestingly, the presence of the graph also influenced the use of other information sources: participants made more use of the statement and evaluated the other data sources more. 
When the graph was particularly helpful, participants made a more focused (i.e., selecting less information from), but at the same time more efficient use (i.e., building mental models) of the other data sources. 
Moreover, with a graph, participants evaluated their own knowledge as being higher, confirming a picture biasing effect. 
When the graph was particularly helpful, though, they reflected more on their own knowledge. 
Finally, there was a systematic increase of silence in the multimedia condition (H$_{3c}$), suggesting that students use more mental effort when solving problems that contain graphs.

\subsection*{Beneficial or biasing picture effect?}
The graphs we used in the current study were designed to fulfill an interpretational function, that is, to represent complex information presented in text or formulae pictorially, and thereby support students' problem solving processes \cite{LeAnCa87}. We therefore expected to find an overall beneficial effect of adding graphs to problems, but no such effect was present in our data. From a theoretical point of view, the stimuli used in this study were designed in line with the \textit{temporal contiguity principle}, that is, that pictorial and the textual material were presented at the same time. However, it is not fully in line with the \textit{spatial contiguity principle} \cite{Ma05a} (also known as the \textit{split attention effect}, \citeNP{ChSw92}). The graph and the explanatory input were given on different parts of the screen and hence might have caused unnecessary visual search of related information, and therefore the absence of a multimedia effect. Split attention may have resulted in that the students invested more mental effort into integrating different parts of the problem, as suggested by the higher proportion of silence in the multimedia condition. 

An alternative explanation for not having found a multimedia effect is that participants did not process the textual information, in particular the formula, in the phonological channel. In this way, they would have bypassed the benefits of the dual-processing assumption in working memory. However, post-hoc inspections of the eye-tracking recordings accompanied by the verbal reports of the participants, showed that a vast majority of the participants verbally described what the mathematical formulas contained; many even read the formulas out loud. Consequently, it is likely that most participants indeed processed the textual information phonologically. Nevertheless, future research should explore when and under which circumstances textual information is actually processed phonologically.

Our results suggest that when seeing a graph, students are more likely to believe in the correctness of the accompanying statements. Students may recognize parts of the input and the graph, and parts (maybe only keywords) of the problem statement and they then say something like “yes this is [\textit{for example}] the triangle inequality, so this must be true”. These results are in line with \citeA{mccabe2008seeing}, who found that including brain images in an article increased the scientific credibility of the results. 
They argue that this may be because the brain images ``provide a physical basis for abstract cognitive processes''. In this study, the graphs rather provide concrete physical interpretations of abstract mathematical formulae. Still, the graphs seem to have a similar persuasive power to affect whether a statement is believed or not.

\subsection*{Processes underlying text-picture integration}
An important aspect of processing multimedia material 
 is to select and integrate information relevant for the task \cite{Ma05a}. We used eye tracking to investigate how information was visually selected, i.e., where the students looked, for how long, and how information was integrated, that is, how often they transitioned between different problem areas. 
When the graph was present, students spent about 20\% of their time looking at it. As a result, they paid proportionally less attention to the input and the to-be-confirmed or rejected problem statement. 
The proportion of time looking at the graph was not related to performance. Interestingly, the more students looked at the problem statement, and the less they looked at the input, the better they performed. 
Furthermore, the more students switched their attention between the problem statement and the graph, the better they performed. 
Thus, the mere presence of a graph that is related to the input is not necessarily helpful. Instead, the graph needs to be integrated with the to-be-confirmed or rejected statement.

Analyses of verbal data revealed that in the multimedia condition participants were often more silent in comparison to the control condition without graphs. As silent pauses are indicators of increased mental effort \cite{YiCh07,Jarodzka2015}, adding graphs to these problems could have increased the amount of mental effort for students. One explanation to this is the fact that the amount of elements in the task increased (i.e., the intrinsic load). A qualitative analysis of two contrasting problems revealed that in the problem where the graph was beneficial it provided students with a representation that was helpful to solve the problem. In the problem where it was most harmful, the graph itself was correct, but the problem statement was not. Still, the graph convinced the students to confirm the statement.

\subsection*{Implications for theory and educational practice}

As a practical consequence, we can conclude that when including graphs in textbooks, it should be ensured that students first and foremost  know exactly what their task is (here: confirm or reject problem statements) to know how to use these graphs. Next, they should always ensure to keep the task itself in mind by integrating the task formulation and the graph. Thus, when designing textbooks, it could be important to consider these integration processes. Future work should investigate different way to facilitate integration by e.g., referring to the graph in the statement and maybe even back from the graph to the problem statement.

Furthermore, implications can be also drawn for theory. Mayer's (\citeNP{Ma05a}) theory of processing information of multimedia clearly describes an optimal scenario, where students actively process all given information, by selecting the relevant information, organizing and integrating it. However, in line with other research \cite<e.g.,>{HoHoHo09}, our study showed that students may simply not take the effort to actively process information and instead use a rather shallow processing strategy (e.g., assuming that when the graph is correct, the rest of the task must also be). In that, pictures could even support such a shallow and misleading processing. The CTML does acknowledge that this optimal way of processing can be hampered by different layout decisions and has thus formulated several design guidelines. Based on the findings in this paper, we suggest that the influence of a picture bias effect should be considered carefully alongside such guidelines.

\subsection*{Limitations and conclusions}
It is evident from discussions we had with students after the test, that the experiment does not  precisely reflect how they normally work with problems of this type at home, in the classroom, or at examinations. 
First, the time to solve a problem was limited and rather short. Such time pressure may lead to more shallow information processing, and therefore a greater picture bias.
Second, they were not allowed to use pen and paper to scribble formulas and figures to organize their problem solving processes.
Finally, these students are typically not exposed to problems where statements need to be falsified, in particular when the information is not presented in their native language.  The implications of using this rather uncommon format for providing the answer need further investigation.
This makes it challenging to construct suitable problems and graphs for these types of studies. 
Nevertheless, the format of the test is still common in other domains. 

From the eye-tracking data and the verbal reports, examples of deep processing, such as building a rich mental model, of the information included in the problems were observed. 
However, the current data analysis does not allow for concrete evidence. Future research should investigate this issue in a qualitative manner.

In summary, graphs were not found to be beneficial per se in the experiment. Only when they were carefully framed and integrated with the problem statement they had a beneficial effect on performance. Otherwise, when the graphs were correct by themselves, they mislead the students to trust the problem statements. Either way, the graphs produced an increase in mental effort. Before including graphs in mathematical texts, teachers and textbook designers should very carefully consider their function and how they integrate with other parts of the information in the problem.

\section*{Acknowledgments}
Financial support from Sveriges l{\"a}romedelsf{\"o}rfattares f{\"o}rbund (SLFF) is gratefully acknowledged.
We thank Johnny Kvistholm for work with several of the figures used in the experiment.

\bibliographystyle{apacite}
\bibliography{refsVector}

\appendix 
\setcounter{table}{0}
\renewcommand{\thetable}{A\arabic{table}}
\section*{Appendix:}
\label{sec:app}
\begin{sidewaystable}[ht]
\vspace{10cm}
\singlespacing
\footnotesize
	\centering
		\caption{Coding scheme for verbal data.} 
	\renewcommand{\arraystretch}{0.7}
			\begin{tabular}{ p{3cm}   p{3cm}  l  p{9cm}  p{4cm} }
    \hline
		\textbf{Main category} & \textbf{Sub-category} & \textbf{Code}  &\textbf{Description  of code}& \textbf{Exemplary statement}\\ \hline

		\multirow{3}{*}{\parbox{2cm}{Search of information from…}} & Input & S1T& Looking / searching / selecting something
specific in the input or reading the input. Nothing
more is uttered than information given in the
input.& “A vector function R …”
(Reading input)\\ \cline{2-5}
   & Statement & S1S& Looking / searching / selecting something specific in the statement. Nothing more is uttered than information given in the statement. & “Then it follows…” (Reading the statement)\\ \cline{2-5}
   & Graph  & S1G&Looking / searching / selecting something
specific in the graph. Simply describing literally
what is presented in the graph. Nothing more
is uttered than information given in the graph. & “the tangent to the
sphere”\\ \hline

\multirow{3}{*}{\parbox{2cm}{Building of mental
model}} & Verbal (input)  & MMVT& Integrating information within the input.  & “because it is the same
vector”\\ \cline{2-5}
   & Verbal (statement)& MMVS& Integrating information within the statement. &“if R is perpendicular to
the derivative”\\\cline{2-5}
   & Graphical & MMG & Integrating information within the graph.& “or a curve on the
surface”\\ \hline

 Prior knowledge activation & & PK & New things / issues / facts are mentioned that
are not displayed on the computer screen. & “the derivative can be
decomposed into…” \\ \hline

\multirow{3}{*}{\parbox{2cm}{Integration of
information}} & Input \& graph& ITG& Connecting / integrating information given in the input with information given in the graph.& “since the derivative is constant it has to be…”\\ \cline{2-5}
   & Input \& statement &ITS &Connecting / integrating information given in
the input with information given in the
statement. &“the square of the length
is constant, such that…”\\\cline{2-5}
   & Graph \& statement & IGS & Connecting / integrating information given in
the graph with information given in the
statement.
& “the radius is
perpendicular to…”\\ \hline

\multirow{4}{*}{\parbox{2cm}{Problem solution }} & & EPS& Uttering, whether the entire statement “Q” is true or false.& “so it is true”\\ \cline{2-5}
   & Input &TS& Uttering, whether something about the input is
true or false.&
“that is natural”\\\cline{2-5}
   & Statement& SS &Uttering, whether something about the
statement is true or false.&
“it seems to be a correct
formula”\\\cline{2-5}
   & Graph & GS& Uttering, whether something about the graph
is true or false.& “so it has to be…”\\ \hline

\multirow{3}{*}{\parbox{2cm}{Metacognitive
statements (‘thinking
about one’s own
thinking’)}} & Evaluating difficulty of task (Positive)& MCT+ &Expressing judgment about high complexity of
the task&
“this is difficult”\\ \cline{2-5}
    & Evaluating difficulty of task (Negative)&  MCT- &Expressing judgment about low complexity of
the task&
“this was not strange at
all”\\\cline{2-5}
   & Evaluating own knowledge (Positive)& MSO+ &Expressing positive judgment about own
abilities and knowledge or about correctness
of own task approach.&
“now I get it”\\\cline{2-5}
& Evaluating own knowledge  (Negative)& MSO- &Expressing judgment about lack of one’s own
knowledge and skills or about mistakes made.&
“I have no idea”\\\cline{2-5}
& Goal orientation &GO &Uttering the goal of the task. &“then the question is…”\\ \hline

Off-topic / not
understandable & & OT & Anything that does not fit the coding schema
or is not understandable.&
“large graph!” \\ \hline

		\end{tabular}
	\label{tab:codingScheme}
\end{sidewaystable}

\end{document}